\documentclass[conference]{IEEEtran}
\IEEEoverridecommandlockouts
\usepackage{cite}
\usepackage{amsmath,amssymb,amsfonts}
\usepackage{algorithmic}
\usepackage{graphicx}
\usepackage{textcomp}
\usepackage{xcolor}
\makeatletter
\newcommand{\linebreakand}{%
  \end{@IEEEauthorhalign}
  \hfill\mbox{}\par
  \mbox{}\hfill\begin{@IEEEauthorhalign}
}
\makeatother
\def\BibTeX{{\rm B\kern-.05em{\sc i\kern-.025em b}\kern-.08em
    T\kern-.1667em\lower.7ex\hbox{E}\kern-.125emX}}

\DeclareUnicodeCharacter{03C3}{$\sigma$}

\begin{document}
\title{Performance and scaling of the LFRic weather and climate model on different generations of HPE Cray EX supercomputers
}

\author{\IEEEauthorblockN{J. Mark Bull}
\IEEEauthorblockA{\textit{EPCC, The University of Edinburgh} \\
Edinburgh, UK \\
m.bull@epcc.ed.ac.uk}
\and
\IEEEauthorblockN{Andrew Coughtrie}
\IEEEauthorblockA{\textit{Met Office} \\
Exeter, UK \\
andrew.coughtrie@metoffice.gov.uk}
\and
\IEEEauthorblockN{Deva Deeptimahanti}
\IEEEauthorblockA{\textit{Pawsey Supercomputing Research Centre} \\
Perth, Australia \\
Deva.Deeptimahanti@csiro.au}
\linebreakand
\IEEEauthorblockN{Mark Hedley}
\IEEEauthorblockA{\textit{Met Office} \\
Exeter, UK \\
mark.hedley@metoffice.gov.uk}
\and
\IEEEauthorblockN{Caoimh\'{i}n Laoide-Kemp}
\IEEEauthorblockA{\textit{EPCC, The University of Edinburgh} \\
Edinburgh, UK \\
C.Laoide-Kemp@epcc.ed.ac.uk}
\and
\IEEEauthorblockN{Christopher Maynard}
\IEEEauthorblockA{\textit{Met Office} \\
Exeter, UK \\
christopher.maynard@metoffice.gov.uk}
\linebreakand
\IEEEauthorblockN{Harry Shepherd}
\IEEEauthorblockA{\textit{Met Office} \\
Exeter, UK \\
harry.shepherd@metoffice.gov.uk}
\and
\IEEEauthorblockN{Sebastiaan Van De Bund}
\IEEEauthorblockA{\textit{EPCC, The University of Edinburgh} \\
Edinburgh, UK \\
S.VanDeBund@epcc.ed.ac.uk}
\linebreakand
\IEEEauthorblockN{Mich\`{e}le Weiland}
\IEEEauthorblockA{\textit{EPCC, The University of Edinburgh} \\
Edinburgh, UK \\
m.weiland@epcc.ed.ac.uk}
\and
\IEEEauthorblockN{Benjamin Went}
\IEEEauthorblockA{\textit{Met Office} \\
Exeter, UK \\
benjamin.went@metoffice.gov.uk}
}

\maketitle

\begin{abstract}
This study presents scaling results and a performance analysis across different supercomputers and compilers for the Met Office weather and climate model, LFRic. The model is shown to scale to large numbers of nodes which meets the design criteria, that of exploitation of parallelism to achieve good scaling. The model is written in a Domain-Specific Language, embedded in modern Fortran and uses a Domain-Specific Compiler, PSyclone, to generate the parallel code. The performance analysis shows the effect of choice of algorithm, such as redundant computation and scaling with OpenMP threads. The analysis can be used to motivate a discussion of future work to improve the OpenMP performance of other parts of the code. Finally, an analysis of the performance tuning of the I/O server, XIOS is presented.
\end{abstract}


\section{Introduction}
Understanding performance and scaling of applications on current supercomputers is important information when considering how to scale applications to Exascale systems. Comparing application performance across multiple systems can provide helpful insight. Three current machines, different generations of HPE Cray EX systems and XC40 systems, are considered here. The first is ARCHER2, the UK National High Performance Computing service hosted by EPCC, the supercomputing centre at the University of Edinburgh. It has 5860 nodes of dual socket, 64-core AMD EPYC ``Rome'' processors. They are connected by the HPE Cray Slingshot (version 10) network. The second machine is Setonix, hosted by the Pawsey Supercomputing Research Centre. It has 1592 nodes of dual socket, 64-core AMD EPYC ``Milan'' processors. These are connected by the HPE Cray Slingshot (version 11) network~\cite{slingshot}. The third machine is the Met Office Cray XC40. It has dual socket, 18-core Intel Xeon ``Broadwell'' processors and the Aries networks.

Weather and climate applications are significant users of HPC resources. Moreover, the predictive power of both Numerical Weather Prediction (NWP) and climate models are limited by the availability of HPC resources and the applications ability to exploit them. Thus, understanding the performance of such models on current HPC architectures can help prepare the applications for Exascale systems. The Met Office is developing a new weather and climate model, named LFRic~\cite{LFRic}, part of the ~Unified Earth Environmental Prediction Framework - Momentum{\textsuperscript{\textregistered}}. Both time to solution and energy efficiency are important for weather and climate applications. Moreover, for a data-intensive application the performance of the I/O system is an important factor in both these metrics. The XML Input Output Server (XIOS)~\cite{XIOS} provides LFRic with a server-client file interaction capacity for HDF5 encoded netCDF files.

In Section~\ref{LFRic}, the model and its components, including the I/O server library XIOS, are described. Section~\ref{perf} contains the performance results and scaling analysis of the dynamical core, GungHo, running at up to approximately 10km global resolution (the current Met Office operational resolution) and beyond on several hundred nodes. Both strong and weak scaling of the model are presented. Comparisons between different compilers, architectures and domain sizes are made. In particular, the scaling of model components for different numbers of OpenMP threads and MPI ranks at fixed resource is examined with different compilers. 

In Section~\ref{IOperf} an analysis of the sensitivity of XIOS I/O performance to XIOS server and output file configuration is presented.  Significant I/O performance optimisations are explored for LFRic running on the Met Office XC40. 


\section{The LFRic weather and climate modelling system}
\label{LFRic}
The dynamical core of the LFRic model, known as GungHo~\cite{GungHo}, considers the fluid dynamics of the atmosphere, and has been developed with both scientific accuracy and computational performance and scaling as considerations. The geometry of the domain is a thin layer of fluid on the surface of a rotating sphere which is vertically stratified due to gravity. It is thus natural to treat the vertical and horizontal degrees of freedom separately. A cubed sphere mesh, treated as an unstructured mesh across the horizontal degrees of freedom, enables scalable communication patterns. The vertical degrees of freedom are treated as a structured mesh, which allows for direct addressing in the vertical dimension in order to to amortise the cost of indirect addressing across the horizontal dimensions~\cite{LFRic}. Finally, a mixed-finite element scheme for the dynamics equations allows for good accuracy of the discretisation. For CPU architectures, data parallelism is employed across the horizontal domain. 

The full atmosphere model (LFRic) employs several physical parameterisation schemes for processes which are not present in the dynamics ({\em e.g.} radiation) or are not resolved at the scale the grid ({\em e.g.} microphysics). These have a finite difference representation and can in general be computed point-wise across the grid, although there are some dependencies, more often in the vertical dimension. Consequently, the full model has more computation and communication than the dynamics. However, the pattern and thus the scaling is similar. The dynamical core can therefore give a good indication of the performance of the full model on a given computer architecture. 

The code is written in a domain-specific language, embedded in modern Fortran. It uses a domain-specific compiler, called PSyclone~\cite{LFRic,PSyclone}, which has been co-designed with the dynamical core. A separation of concerns approach is used to keep the science/maths code separate from the parallel code. PSyclone is then able to generate the parallel code for both distributed memory parallelism using MPI and shared memory parallelism using OpenMP. To support the use of the unstructured mesh, a library called Yet Another Exchange Tool (YAXT)~\cite{YAXT} is employed to determine the routing tables for the halo-exchange communication pattern on the unstructured mesh.

The application can be run with a choice of using redundant computation to set or compute the values of the halos. This computation is redundant as neighbouring MPI ranks both compute the same values. There is then no need to communicate data to update the halos, resulting in less communication.

\section{Performance of GungHo}
\label{perf}
In this section we report on the performance of GungHo, the dynamical core of LFRic, on both HPE Cray systems. In all cases, nodes are fully populated with processes/threads running on all 128 cores per node. We use three different sizes of the cubed sphere mesh: C256, C512 and C1024, all with 120 vertical levels. The total number of horizontal grid points is the square of the mesh number multiplied by 6 (the number of faces of the cube): so, for example, a C512 mesh has $6 \times 512^2 = 1,572,864$ horizontal mesh points. We refer the number of horizontal grid points per core as the \textit{local area}. In practice, the model timestep needs to satisfy a weak Courant–Friedrichs–Lewy condition, such that halving the model grid spacing (e.g. from C512 to C1024) also requires a halving of the model timestep. However, for the purposes of this study,  we simply report the execution time per timestep, rather than computing the model performance in, for example, simulated seconds per second. In each case we run the model for 96 timesteps, disable I/O, and do not time the initialisation and finalisation phases. 

\subsection{Hardware and software environments}
We have conducted our performance experiments on two HPE Cray XE systems, ARCHER2 and Setonix. Table \ref{tab:systems} shows the important characteristics of the hardware and software environments for the two systems and how they differ. 
\begin{table*}[h!]
    \centering
    \begin{tabular}{|l|c|c|c|}
    \hline
     \textbf{System}    & \textbf{ARCHER2} & \textbf{Setonix} &\textbf{XC40}\\ 
     \hline
      CPU model    & AMD EPYC 7742 ``Rome'' 64 cores & AMD EPYC 7763 ``Milan'' 64 cores & Intel Xeon E5 2695 v4 \\
     CPU clock frequency    & 2.0GHz (capped)  & 2.45GHz & 2.1GHz\\
    CPUs per node   & 2 & 2 & 2\\
    No. of nodes  & 5600  & 1600 & 2000(+) \\
    Level 3 cache per CPU    & 16 $\times$ 16MB & 8 $\times$ 32MB & - \\
    NUMA domains per CPU & 4 & 4 & 1\\
    Interconnect     & Slingshot 10  & Slingshot 11 & Aries \\
      Cray MPICH version   & 8.1.23 & 8.1.25 & 7.3.1\\
     Cray Fortran compiler version    & 15.0.0 & 15.0.1 & - \\
    GNU Fortran compiler version    & 11.2.0 & 12.2.0 & - \\
    \hline 
    \end{tabular}
    \vspace{0.25cm}
    \caption{Hardware and software characteristics of the HPE Cray EX systems ARCHER2 and Setonix and the Met Office Cray XC40. For the latter case the Intel17 compiler was used.}
    \label{tab:systems}
\end{table*}
The CrayPAT profiling tool is used to measure the performance of the application. As well as identifying the most time-consuming routines, the profiles allow us to break down the execution time into communication and computation times. This split is straightforward and unambiguous for this code, since it is written in master-only style (all MPI calls are outside of OpenMP parallel regions, and all communications are blocking: there is little scope for overlapping in the algorithms used). 

\subsection{Results}
\begin{figure}
    \centering
    \includegraphics[width=1.0\columnwidth]{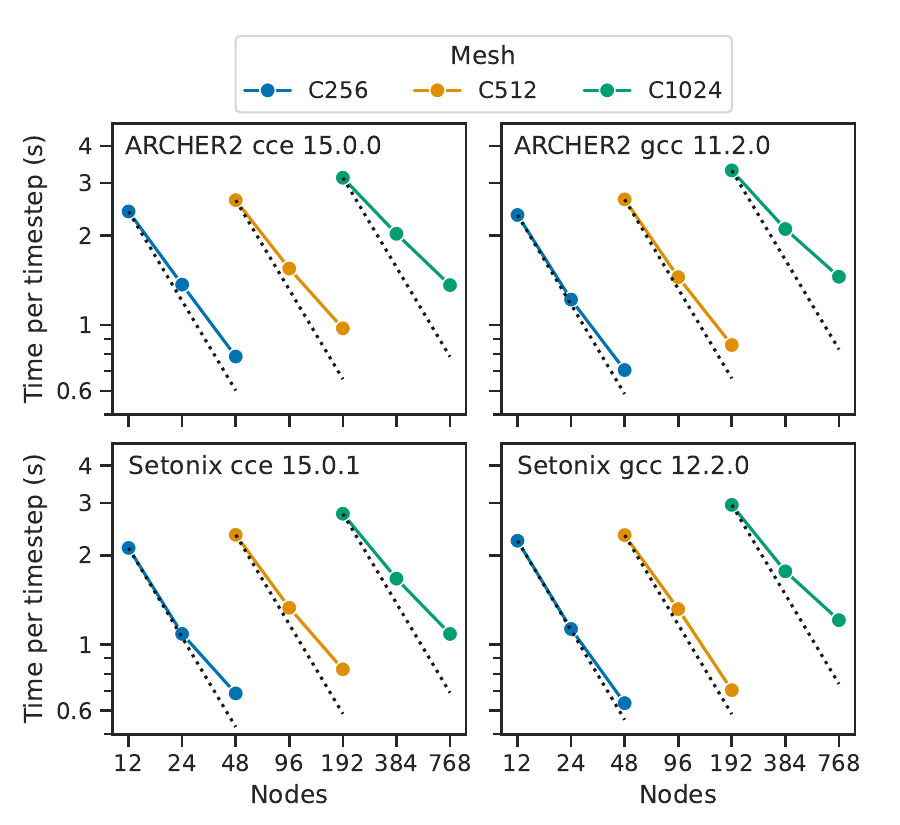}
    \caption{Strong scaling behaviour of GungHo for different mesh sizes on ARCHER2 (top row) and Setonix (bottom row) using Cray compilers (left column) and GNU compilers (right column).}
    \label{fig:gungho-loglog}
\end{figure}
\subsubsection{Strong scaling}
Figure~\ref{fig:gungho-loglog} shows the strong scaling behaviour of GungHo on the two systems, ARCHER2 and Setonix, with both Cray and GNU compilers. The figure shows the execution time (in seconds) per model timestep against the number of MPI ranks for different mesh sizes (C256, C512 and C1024). In all cases, we run 32 MPI ranks per node, each with four OpenMP threads. Due to memory requirements, it is not possible to perform the experiments across a wide range of rank counts for a given mesh size, so there are a limited number of points in each scaling curve. The principal trends that can be observed are that the strong scaling deviates more from the ideal (dotted lines) for larger mesh sizes, and that the deviation is greater for all mesh sizes on ARCHER2 compared to Setonix.  We will examine the differences between systems and compilers in more detail below.  

\subsubsection{Weak scaling and optimal thread count}
\begin{figure*}
    \centering
    \includegraphics[width=0.95\textwidth]{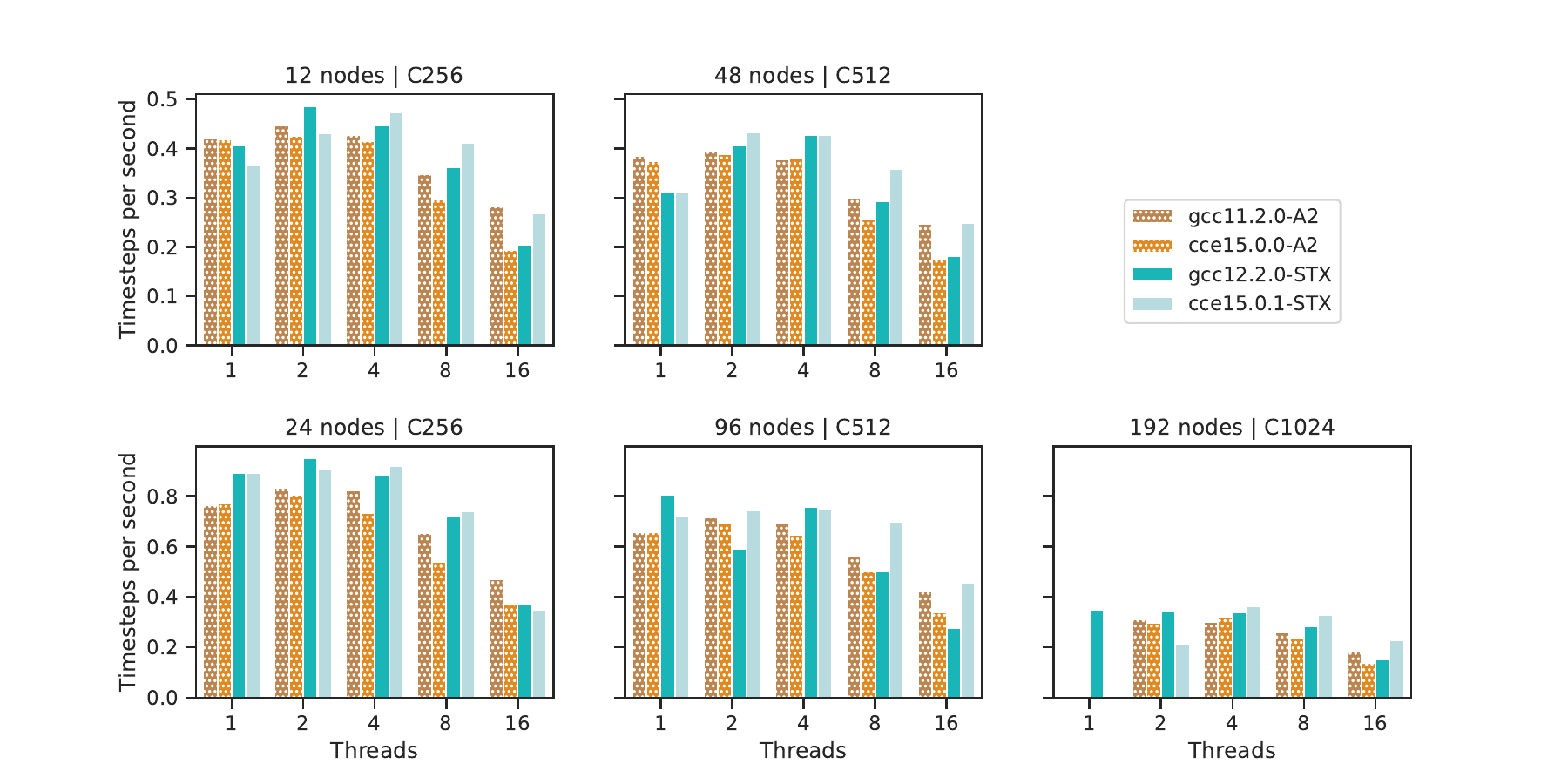}
    \caption{Weak scaling and threading performance of GungHo for a local areas of size 256 grid cells per core (top row) and 128 grid cells per core (bottom row).}
    \label{fig:gungho-weak-threads}
\end{figure*}

Figure~\ref{fig:gungho-weak-threads} shows the performance in timesteps per second (high is good) for different mesh sizes and node counts. In each row of the figure the local area is kept constant: in the top row both configurations have 256 horizontal grid points per core, while in the bottom row they all have 128 horizontal grid points per core. For each mesh size/node count combination, we also vary the number of OpenMP threads and MPI ranks per node, subject to the product of these two values always being 128, the number of cores per node. Some of the C1024 mesh runs on 192 nodes with one OpenMP thread per rank failed due to insufficient memory. 

We can observe that the weak scaling behaviour is reasonably good. The performance does not decrease much as we increase the mesh size and number of nodes, until we reach the C1024 on 192 nodes case. We can also see that the best performance is obtained by running either one, two or four OpenMP threads per MPI rank, depending on the configuration, but eight or sixteen threads per rank is never the best choice. The reasons for this are examined in more detail in Section \ref{commvcalc}. In general the differences in performance between the systems and compilers are small, though the best performance for any of the configurations is always obtained on Setonix, and not on ARCHER2. Section \ref{sysandcomp} will focus more closely on these differences.

\subsubsection{Communication versus computation}
\label{commvcalc}
We now consider the breakdown of execution time into communication and computation, as reported by the CrayPAT profiling tool. On ARCHER2, we can further divide the communication time into halo exchanges (MPI\_ISend, MPI\_Irecv and MPI\_Waitall calls in the YAXT library) and collective communication (predominantly MPI\_Allreduce calls require for global sums in the semi-implicit solver). Unfortunately, this breakdown is not obtainable for Setonix, as CrayPAT apportions time lower down in the MPI call tree. CrayPAT also breaks down computation in user code (USER) and system and runtime library calls (ETC). Profiles on Setonix attribute small amounts of time in different groups not present for ARCHER2. Thus, for Setonix, in order to obtain comparable profiles, time reported in FABRIC is apportioned to MPI and time in PTHREAD to ETC. We also show the percentage of time that CrayPAT does not account for (the reported percentages of samples do not quite add up to 100\%).  

\begin{figure}
    \centering
    \includegraphics[width=0.95\columnwidth]{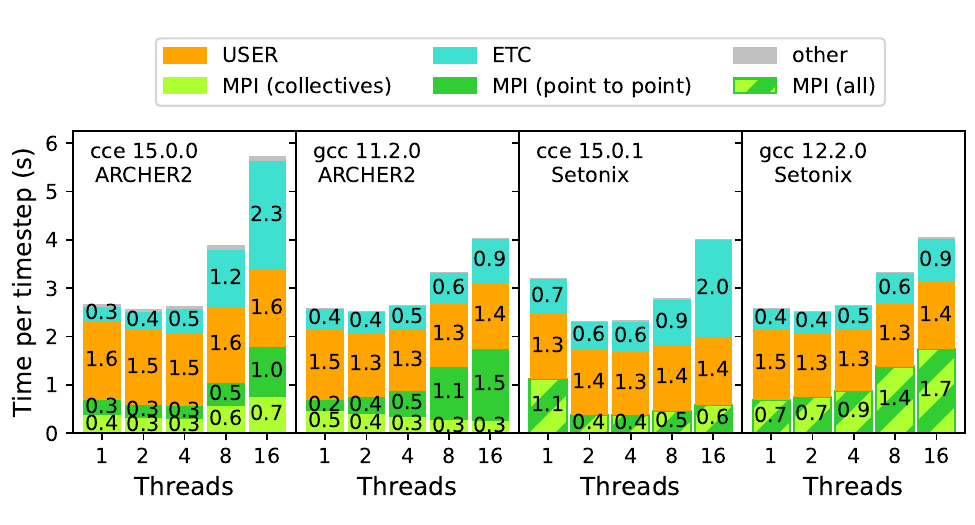}
    \caption{Breakdown of execution time of GungHo for a C512 mesh on 48 nodes, on ARCHER2 (left) and Setonix (right) using the Cray compiler and GNU compiler.}
    \label{fig:gungho-profile-threads}
\end{figure}

Figure~\ref{fig:gungho-profile-threads} shows the breakdown of execution time of GungHo on ARCHER2 and Setonix for a C512 mesh on 48 nodes (this corresponds to the centre plot on the top row of Figure~\ref{fig:gungho-weak-threads}, but here we are showing execution time per timestep rather than performance in timesteps per second). We can observe that the generally small benefit (i.e. lower execution time) of using two or four threads per MPI rank comes from small reductions in user code time and MPI collective time. These can be attributed respectively to a reduction in the amount of redundant computation in halos, and to fewer MPI processes being involved in the global sums. The benefit on using more than one thread is larger on Setonix with the Cray compiler, but is not clear why the MPI time is so much longer with one thread.

As the number of threads per MPI rank is increased to 8 and 16, the execution time increases significantly. While the USER time remains approximately constant, the MPI point-to-point time increases. With fewer MPI ranks, despite the fact that the total amount of halo data decreases, the size of the halos increases, and fewer cores are involved in the halo exchange. For example, as the number of threads per MPI rank increases from 4 to 16, the total amount of halo data reduces by a factor of two, but the individual halos are doubled in size, and the number of MPI ranks involved in the exchange is reduced by a factor of four. The net result is that the halo exchanges take longer as the number of threads per process increases. 

For the collective communication, using the GNU compiler, the time taken continues to reduce as the number of threads per MPI rank is increased above 4, as we might expect, since the number of MPI ranks involved in the global sum continues to reduce. However, the opposite is true for the Cray compiler, and the reason for this is unclear. It is also interesting to note that the total time spent in MPI is larger for the GNU compiler, but for one, two and four threads per rank, this is (co-incidentally) compensated for by a lower USER time, resulting in very similar total execution times for the two compilers. Since the MPI library version is the same, the differences must be result of building the library with the two different compilers. 

Also of note is the increase in ETC time as the number of threads per rank increases. For the GNU compiler this can be attributed to the cost of the barriers used to synchronise threads at the end of OpenMP parallel loops. This additional barrier cost is also present for the Cray compiler on both systems, but the larger ETC times also include POSIX mutex lock and unlock calls. We have not so far been able to determine the source of these, but memory allocation/deallocation is one possibility. 

\begin{figure}
    \centering
    \includegraphics[width=\columnwidth]{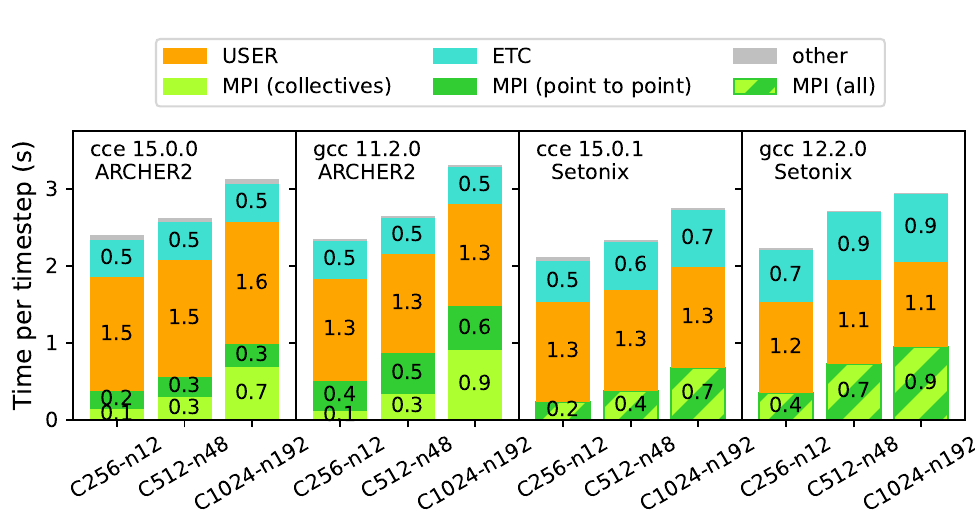}
    \caption{Breakdown of execution time of GungHo on ARCHER2 for a fixed local area of 256 grid cells, on ARCHER2 (left) and Setonix (right) using the Cray compiler and GNU compiler.}
    \label{fig:gungho-profile-weak}
\end{figure}
Figure~\ref{fig:gungho-profile-weak} shows the same breakdown of execution time, for mesh size and node counts configurations that all have the same local area (256 grid cells per core), and four threads per MPI rank. Here we can clearly see that on ARCHER2 the loss of weak scaling (i.e. the increase in execution time as the grid size and node count increase) is almost all due to an increase in MPI collective communication time. This is a result of the increase in the number of MPI ranks involved in the global sums. The computation time remains constant, and there is a small increase in the MPI point-to-point (halo exchange) time. On Setonix, it would be reasonable to assume that the same increase in collective time is also responsible for the loss of scaling. Execution times on Setonix are a little lower than on ARCHER2. On Setonix CrayPAT records less MPI time than on ARCHER2 but more USER time, but this could be a profiling artefact. 

\subsubsection{Comparison of systems and compilers}
\label{sysandcomp}
To better visualise the differences between systems and between compilers, we compute ratios of execution times. 
\begin{figure*}
    \centering
    \includegraphics[width=0.95\textwidth]{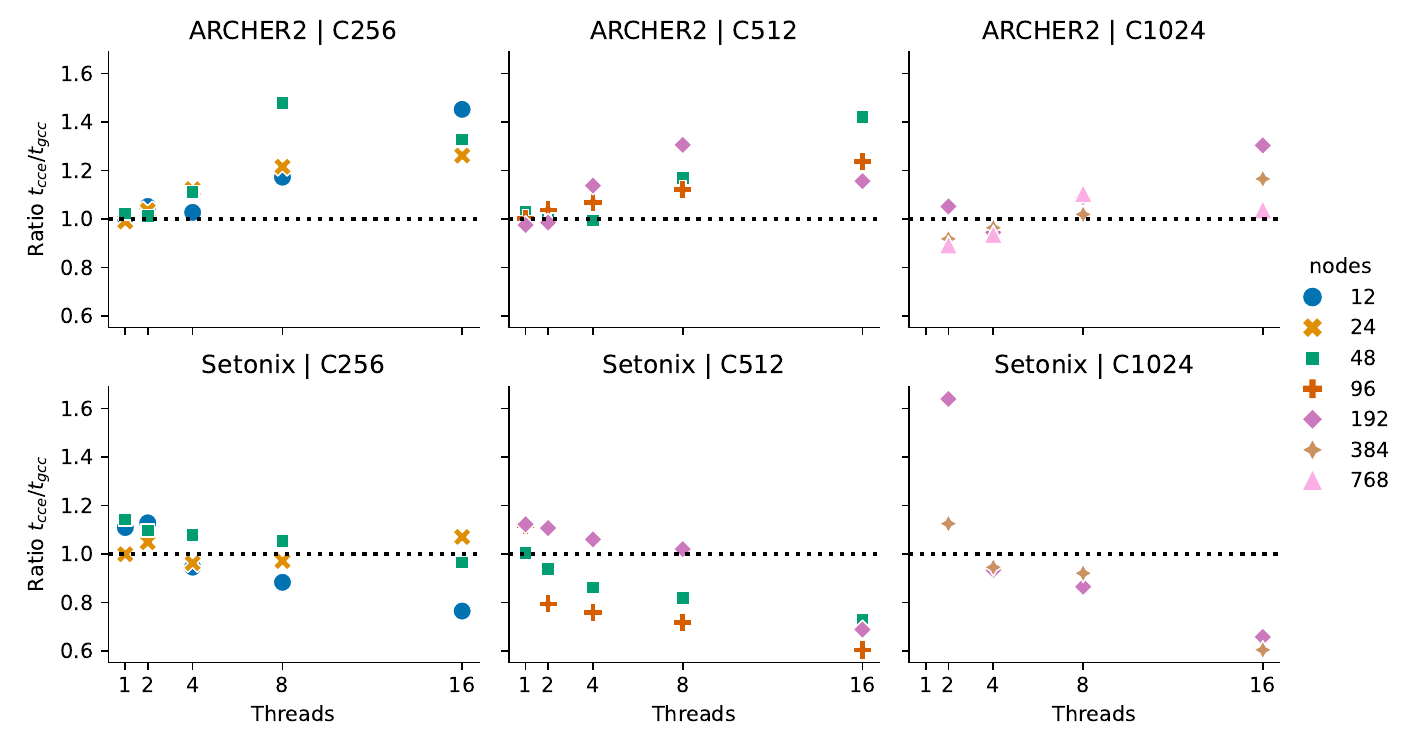}
    \caption{Ratio of execution time of GungHo using the Cray compiler to that using the GNU compiler on ARCHER2 (top row) and Setonix (bottom row).}
    \label{fig:gungho-ratio-compilers}
\end{figure*}
Figure~\ref{fig:gungho-ratio-compilers} shows the ratio of execution time of GungHo using the Cray compiler to that using the GNU compiler (so values larger than one indicate that the GNU compiler has better performance). On ARCHER2, the performance with the GNU compiler is equal to, or better than, the Cray compiler in most cases, and the difference tends to become larger with higher numbers of threads per MPI rank. This is likely due to the mutex lock/unlock overheads identified in Section \ref{commvcalc}. On Setonix, we observe the opposite trend: GNU is faster for low numbers of threads per rank. However, for the optimal thread numbers (1, 2 or 4), the differences between the two compilers are quite small. 

\begin{figure*}
    \centering
    \includegraphics[width=0.95\textwidth]{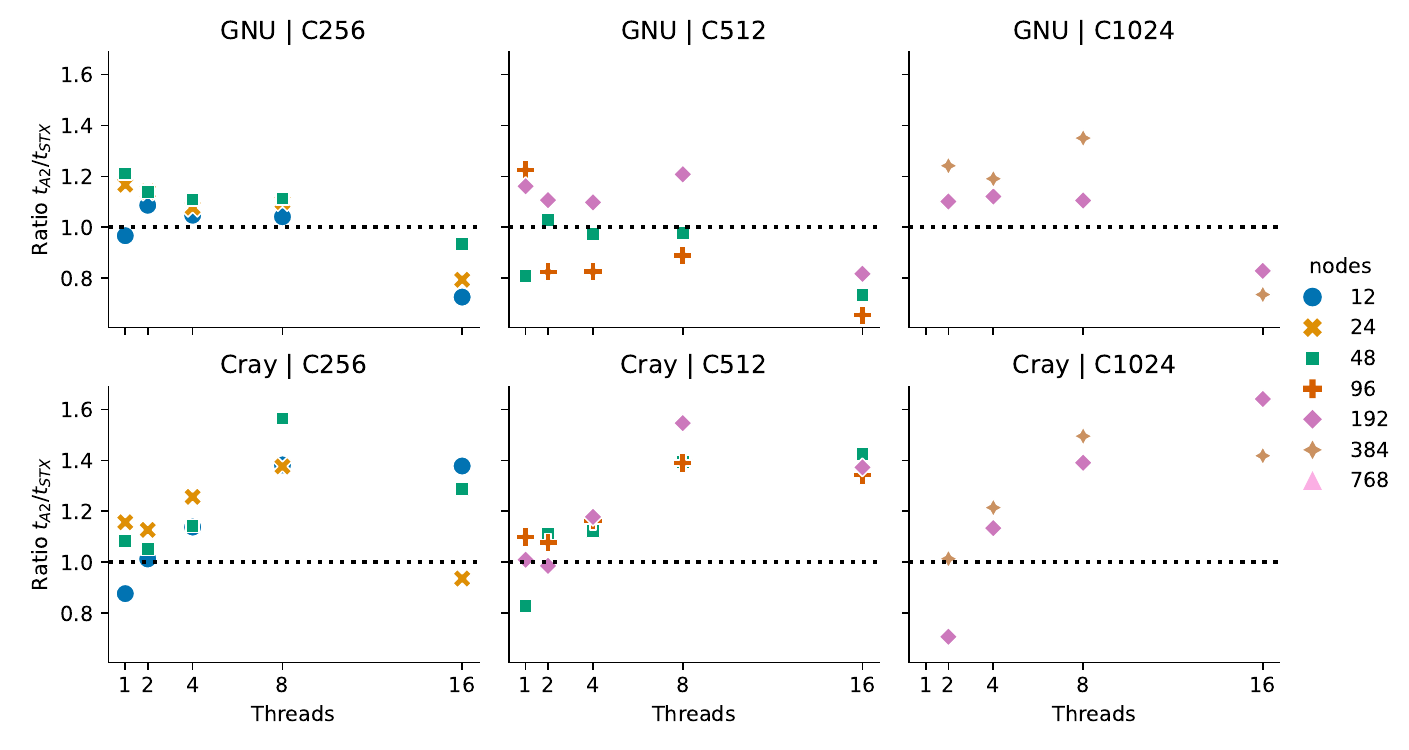}
    \caption{Ratio of execution time of GungHo on ARCHER2 compiler to that on Setonix,  using the GNU compiler (top row) and Cray compiler (bottom row).}
    \label{fig:gungho-ratio-machine}
\end{figure*}
Figure~\ref{fig:gungho-ratio-machine} shows the ratio of execution time of GungHo on ARCHER2 to that on Setonix (so values larger than one indicate that Setonix has better performance).
This shows a rather mixed picture: in the majority of cases Setonix performs better, but there are few clear trends. With the Cray compiler, the difference in performance between the two systems is largest (for most mesh/node configurations) at 8 threads per rank. This could be because the L3 caches on Setonix are shared between 8 cores, compared to 4 on ARCHER2. However, this effect is less marked for the GNU compiler. 

\subsection{Summary}
To summarise the findings in this Section, we can say that GungHo scales acceptably well up to 768 nodes (24576 MPI ranks, 98304 cores) for relevant mesh sizes (the initial operational forecasting configuration is expected to be a C896 mesh with around 70 levels running on 147 nodes). We have also seen that the main factor limiting scalability is the global sums in the semi-implicit solver. 

We have shown that running with 2 or 4 threads per MPI rank gives some modest performance gains over MPI only, and that, with some variations depending on model configuration and node count, the performance of the Cray and GNU compilers does not differ very much. We also observe slightly better performance on Setonix compared to the older ARCHER2 system overall, but again the differences are generally quite small. 

\section{I/O performance}
\label{IOperf} 

LFRic uses XIOS (XML I/O Server), released and maintained by Institut Pierre-Simon Laplace (IPSL~\cite{XIOS}) to manage file interactions for loading and saving data.  XIOS buffers data and manages parallel reads and writes to netCDF files to hide I/O from simulation processes as much as possible. This limits the impact of I/O on simulation ranks.

To maximise the benefit of using the I/O server, we must ensure that the ratio of time spent by XIOS waiting for a free buffer to the total compute time is close to zero. There are several ways to minimise this ratio, including:
\begin{itemize}
\item Increase the rate of writing data, to free buffers up more quickly.
\item Increase the number of XIOS servers, although this may compromise the above if not done carefully.
\item Increase the XIOS buffer size, although this may increase the memory footprint.
\end{itemize}

\subsection{Measures}

The key measures for analysing I/O performance identified and used here are:

\begin{itemize}
\item Wall clock time (in seconds).
\item XIOS Client buffer wait \%
  \begin{itemize}
  \item The amount of time that XIOS client ranks spend waiting for an available buffer divided by the total time XIOS clients are active.
  \item This gives an indication of how well the XIOS servers are enabling simulation ranks to offload data.
  \end{itemize}
\item XIOS Server write rate (MiB/second)
  \begin{itemize}
  \item The overall write rate of all XIOS server ranks to write all of the data on demand.
  \end{itemize}
\end{itemize}

Performance measurements are quite variable on supercomputer hardware, with many factors influencing I/O performance.  It is important that sufficient I/O load is provided for performance runs, to give a reasonable chance of recognising signals for performance change within the noise of overall platform variation.

\subsection{Processor Numbers and Buffer Sizes for XIOS}

\begin{figure*}
  \centering
  \includegraphics[width=0.9\textwidth]{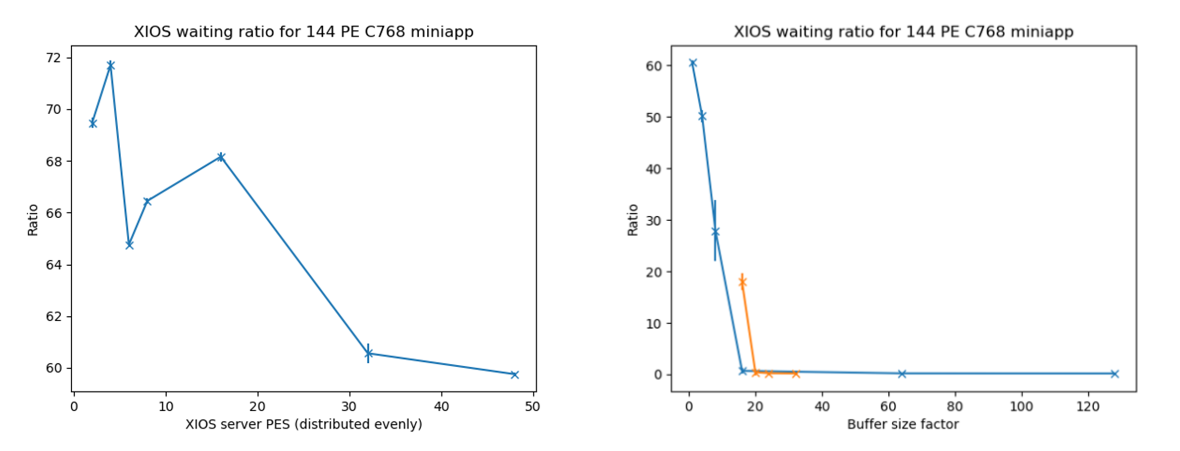}
  \caption{The effect of increasing the number of XIOS servers (L) and buffer size factor (R) on the waiting ratios for the iodev miniapp. Our target is to be as close to zero as possible - as this means data writing is hidden completely from the model. We see that for these two parameters, there is much more impact from increasing the buffer size. On the right hand plot, the orange line represents 16 PEs, and the blue line 32 PEs for XIOS.}
  \label{fig:waiting_ratios_buffer}
\end{figure*}

Using the Met Office XC40s and an \texttt{iodev} miniapp from the LFRic repository trunk, we explore the sensitivity to processor numbers and buffer sizes. These results are shown in Figure~\ref{fig:waiting_ratios_buffer}. We see that increasing the number of XIOS servers only gets us so far in improving performance, but to really minimise the wait ratio we have to keep an eye on the buffer size.

\subsection{C192 Diagnostic Load Tests}

A low resolution C192 configuration was used on the Met Office XC40 to investigate enhanced diagnostic load impacts on I/O performance. The C192 test runs for 48 model hours, and during that runtime writes the following fields at the given output frequencies. The overriding concern is the output frequency rather than the nature of each individual field.
\begin{itemize}
\item 38 fields at 18 hour output
\item 6 fields at 12 hour output
\item 9 fields at 9 hour output
\item 27 fields at 3 hour output
\item 99 fields at 1 hour output
\end{itemize}
In summary, in 48 hours a total of 5329 fields are written, which comes to approximately 400 GiB of data in this case. The job runs with 864 LFRic PEs, and 72 XIOS PEs spread across 4 nodes - 28 nodes total for the job. The total run time is 2857s, with 2632s spent in XIOS processing events (writing data). More of a concern is that each client spends on average 902s during the model run waiting for a data buffer, \emph{time that cannot be spent doing computation}.
\\

\begin{table*}
    \centering
    \begin{tabular}{|l|c|c|}
    \hline
     & \textbf{Baseline} & \textbf{Performance Run} \\ 
     \hline
      \textbf{Characteristics}  &  &  \\
      XIOS Nodes & 22 / 153 (14.4\%) & 24 / 155 (15.4\%) \\
      XIOS Ranks per Node & 17.81 & 19 \\
      Lustre Striping  & None & Full \\
      Log Levels  & XIOS:50, LFRic:Info & XIOS:1, LFRic:Warn \\
      \textbf{Measures}  &  &  \\
      Test Wall Clock Time: & 5140.67 s ±σ 52.12 & 3861.52 s ±σ 21.36 \\
      Data Intensity & 0.15 GB per core hour  & 0.20 GB per core hour \\
      Server Process av. write rate & 377.04 MiB/s ±σ 12.23 & 953.43 MiB/s ±σ 64.87 \\
      Client Time Buffer Wait \%  & 7.56 \% ±σ 1.60 & 1.44 \% ±σ 0.66  \\
      Client time Buffer Wait  & 382.24 s ±σ 81.25 & 53.49 s ±σ 24.63  \\
    \hline 
    \end{tabular}
    \vspace{0.25cm}
    \caption{I/O Performance results comparing a C896 (11km) baseline and a tuned performance scenario.}
    \label{tab:IOPerformance}
\end{table*}

\subsubsection{C192 Initial test}
Considering the optional XIOS configuration of ``level 2'' servers, and an ansatz determined results of detailed sensitivity analysis on XIOS stand alone I/O tests, the following modifications were made: the total number of XIOS servers was reduced to $16$, distributed into $8$ ``level 1'' servers and $8$ ``level 2'' servers, split into $4$ pools of $2$ servers each. We see poor write scaling for large numbers of PEs (smallish pools). We write about 30 files, of varying sizes, so a modest number of pools was chosen, and we need enough memory to process the data (hence remaining at 4 nodes). Using these modifications, the total runtime is reduced to 1257s ($2.3\times$ speedup), and the average time spent waiting for a free buffer is now 42s ($4.5\%$ of the original case), \emph{using the same amount of HPC resource}.
\\
\subsubsection{C192 Further consideration}
\begin{figure}
  \centering
  \includegraphics[width=1.0\columnwidth]{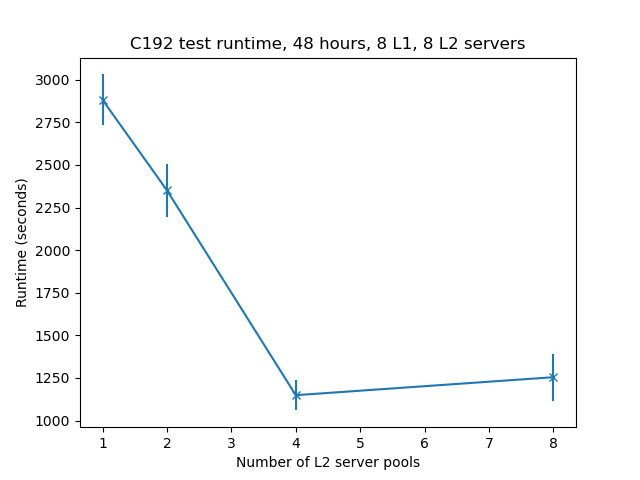}
  \caption{Runtime of the C192 test on 864 LFRic PEs, when using XIOS server across 4 nodes, with 16 individual servers, 8 level 1, 8 level 2. We show the effect of increasing the number of server pools, (and hence decreasing the number of servers per pool) on the total runtime. We see that in this configuration having 4 or 8 servers per pool is not conducive to good performance.}
  \label{fig:server2_c192}
\end{figure}

The server pools are now changed, keeping a $50\%$ ratio between level 1 and level 2 servers, whilst still running across the 4 nodes with 16 servers. In this particular case, increasing the proportion of level 2 servers, or decreasing the node count, leads to ``out of memory'' errors. These results are shown in Figure~\ref{fig:server2_c192}. In this case we run from 1 pool of 8 level 2 servers, to 8 pools of 1 server each. We see a significant sensitivity to the number of pools, and processors per pool, leading to a very significant reduction of runtime.

These experiments also show that simply giving more processors to XIOS does not lead to an increase in performance and we have therefore reduced the number of servers from $72$ to $16$. The increase in performance comes from understanding the sensitivity of XIOS to parallel writing on multiple MPI ranks, along with the ability to further parallelise using files per server pool. It is likely that optimal configurations will be different depending on the exact model configuration, and the total HPC resource that will be used is dependent on the volume of data, as sufficient memory must be available to XIOS.

\subsection{C896 Scenarios}
I/O Performance test results for LFRic were run on Met Office Cray XC40 architecture using Lustre disk based storage and XIOS. The LFRic atmosphere C896 resolution provides a 10km horizontal resolution for the model. A derivative of the LFRic C896 test run was created with enhanced hourly diagnostic load and run with multiple configuration changes (3 runs per scenario) to assess I/O performance sensitivity to configuration factors whilst writing 1.1TiB of diagnostic data.

The baseline uses 153 nodes, 131 simulation nodes (35 ranks per nodes, 4704 ranks), and 22 separate XIOS I/O server nodes (with 17/18 ranks per node). A node is either hosting simulation ranks or XIOS server ranks. The baseline and scenarios would benefit from future analysis, targeting EX architectures, newer XIOS versions and Flash storage. There are positive signals from isolated XIOS performance analyses conducted on ARCHER2 exploring new XIOS developments (that are not yet released) and XIOS level 2 server configurations, separating data gathering nodes and I/O interaction nodes into pools and tuning pool configurations to data. These configurations are not yet included in the presented scenarios, but offer interesting opportunities for further optimisation targeting.

\subsection{C896 Performance Sensitivity Results}

\begin{figure*}
  \centering
  \includegraphics[width=0.9\textwidth]{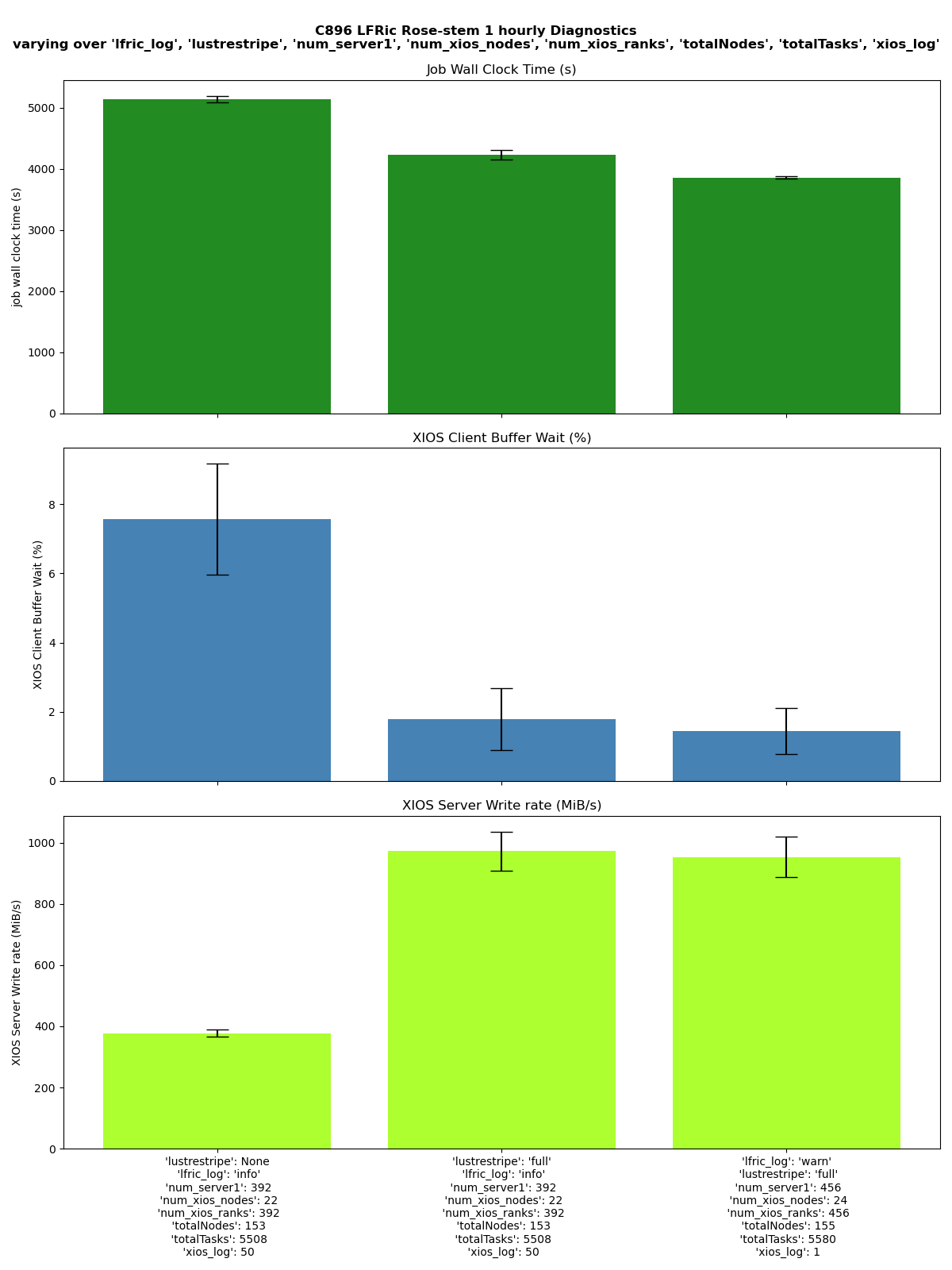}
  \caption{I/O Performance Enhancements from Configuration Tuning.}
  \label{fig:lfricIO}
\end{figure*}

\begin{figure*}
  \centering
  \includegraphics[width=0.9\textwidth]{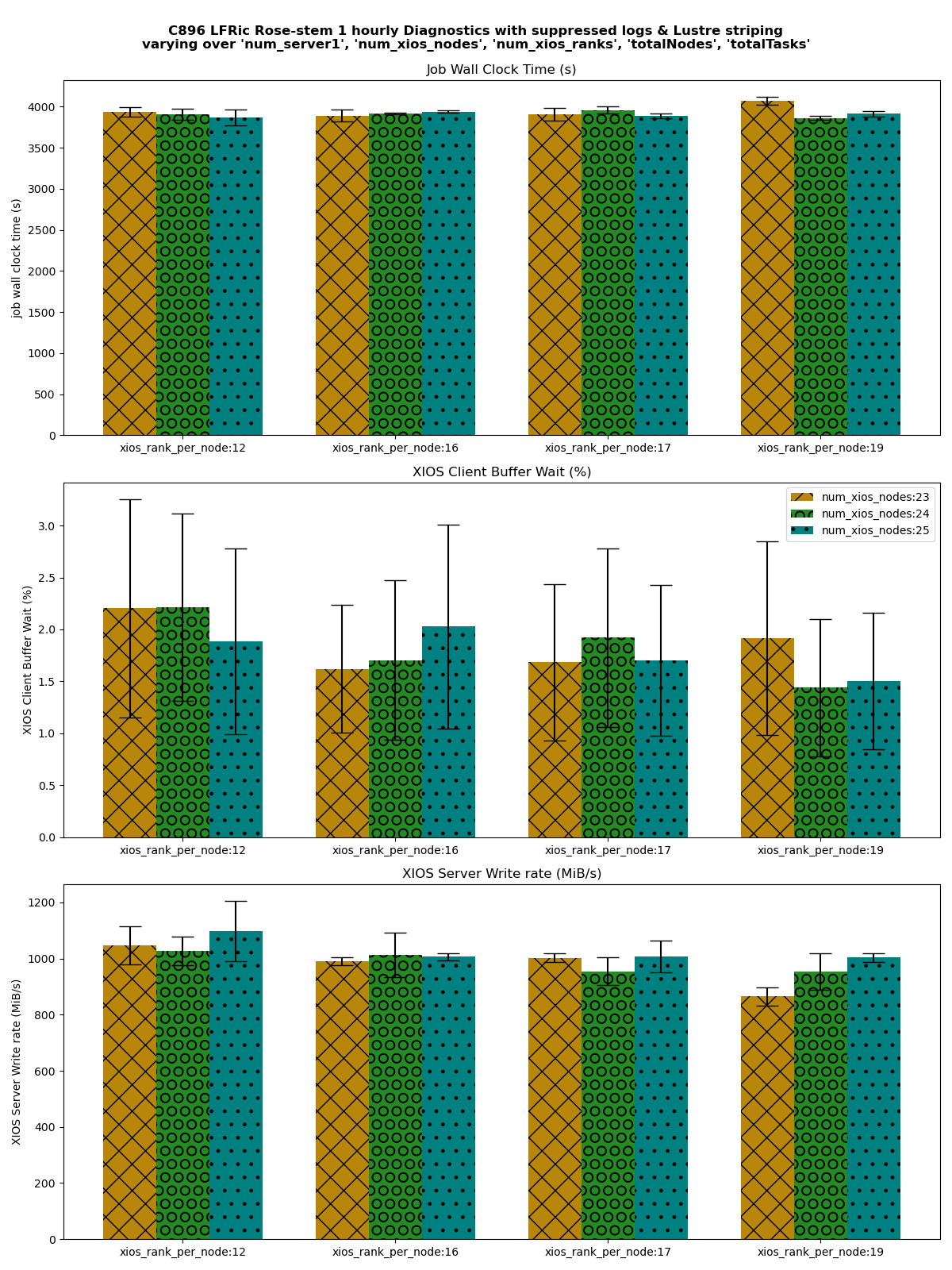}
  \caption{I/O Sensitivity to Number of XIOS Nodes \& Ranks.}
  \label{fig:xiosNodesRanks}
\end{figure*}

By tuning configuration parameters, a $25\%$ improvement in wall clock time, a $253\%$ increase in server write rate and a $5\times$ reduction in client wait percentages were demonstrated for a C896 run writing 1.1TB of data (see Table ~\ref{tab:IOPerformance} for a numerical summary of results). Figure~\ref{fig:lfricIO} compares the wall clock time, XIOS client buffer wait time \% and averaged server write rate for the baseline case, for adding Lustre striping, and for a scenario with Lustre striping, reduced logging and adjusted XIOS nodes and ranks. The error bars show standard deviations for measured quantities, combining variation within a scenario run, and across 3 independent scenario runs. The largest benefit is from Lustre striping -- without this, there is little impact on I/O performance from other factors.  This is consistent with Lustre's documentation on parallel writes to large file objects. Figure~\ref{fig:xiosNodesRanks} demonstrates the sensitivity of I/O performance measures to the choice of number of XIOS nodes and XIOS ranks per node.  The variability within runs and across runs is large compared to the difference in performance measures. Thus, whilst there is some benefit in fine tuning numbers of nodes and ranks for XIOS, the sensitivity of I/O performance to these choices is not large. 

\section{Conclusions and Future Work}
\label{conc}

\subsection{GungHo performance}
GungHo, the dynamical core of LFRic, shows good scalability up to, and beyond the mesh size and node counts that will form the initial operational forecasting configuration. The benefits of the more recent generations of CPU model and network on Setonix are detectable, but quite modest for this part of the application. 

There are, however, ongoing efforts to further improve performance. We are investigating code and data layout restructuring to improve vectorisation of loops over vertical levels. There are opportunities to improve locality and reduce synchronisation overheads by restructuring the OpenMP parallel loops over horizontal grid cells, and also to attempt to realise overlapping of halo exchanges with computation. Performance anomalies uncovered in this study (e.g. time spent in POSIX lock routines using the Cray compiler, and the higher cost of MPI calls using the GNU compiler) will be further investigated.

\subsection{I/O Performance}

I/O is a crucial factor in model performance, with more proportional impact for high diagnostic loads and high horizontal model resolutions. Significant performance improvements can be achieved through minor configuration changes to model interactions with storage systems. XIOS's two level server configurations show real promise for performance gains, but these do add implementation complication. For the C896 I/O scenarios, some issues on ancillary loading are unresolved, so the benefits are not yet as well analysed, however this is a focus for the near future.

XIOS performance sensitivity testing on ARCHER2 has shown significant performance gains and promising tuning opportunities for XIOS, including development versions of XIOSv3, XIOS level 2 servers and use of flash-based storage. These configurations have not yet been combined with the LFRic I/O performance sensitivity analysis that was conducted on the Met Office XC40s with the C896 configuration. Future work will target extending the C896 performance sensitivity analyses to I/O performance runs of LFRic on the Cray EX architecture with further XIOS/LFRic configurations explored.

\section*{Acknowledgment}
This GungHo performance analysis work was undertaken as part of the Met Office Academic Partnership and used the ARCHER2 UK National Supercomputing Service (https://www.archer2.ac.uk). This work was also supported by resources provided by the Pawsey Supercomputing Research Centre with funding from the Australian Government and the Government of Western Australia.
\bibliography{refs}
\bibliographystyle{IEEEtran}

\end{document}